\documentclass[11pt]{elsart}
\usepackage{doublespace}
\usepackage{epsfig}
\begin{document}
\newcommand{\kpp}{K_S\to2\pi^0}
\newcommand{\kppp}{K_S\to3\pi^0}
\newcommand{\eppp}{\eta\to3\pi^0}
\newcommand{\pksl}{\phi\to K_SK_L}
\newcommand{\newlim}{1.4}
\newcommand{\rpp}{(\ref{ks3eq2})}
\newcommand{\rppp}{(\ref{ks3eq1})}
\newcommand{\retg}{(\ref{ks3eq3})}

\begin{frontmatter}
\begin{singlespace}
\title{\bf Upper limit on the $\kppp$ decay}
\author{ M.N.Achasov }
\author{ V.M.Aulchenko }
\author{ A.V.Berdyugin }
\author{ A.V.Bozhenok }
\author{ A.D.Bukin }
\author{ D.A.Bukin\thanksref{addr} }
\author{ S.V.Burdin }
\author{ T.V.Dimova }
\author{ V.P.Druzhinin }
\author{ M.S.Dubrovin }
\author{ I.A.Gaponenko }
\author{ V.B.Golubev }
\author{ V.N.Ivanchenko }
\author{ I.A.Koop }
\author{ A.A.Korol }
\author{ S.V.Koshuba }
\author{ I.N.Nesterenko }
\author{ E.V.Pakhtusova }
\author{ A.A.Salnikov }
\author{ S.I.Serednyakov }
\author{ V.V.Shary }
\author{ Yu.M.Shatunov }
\author{ V.A.Sidorov }
\author{ Z.K.Silagadze }
\author{ Yu.V.Usov }
\thanks[addr]{ e-mail: D.A.Bukin@inp.nsk.su }
\address{ Budker Institute of Nuclear Physics, 630090, Novosibirsk, Russia }
\address{ Novosibirsk State University, 630090, Novosibirsk, Russia }
\date{}
\begin{abstract}
The search for CP-violating $\kppp$ decay was performed with SND detector
at VEPP-2M $e^+e^-$ collider. The total amount of data
corresponding to 7 million produced $K_S$ mesons was analyzed.
No candidate events were found, giving an upper limit of the decay branching
ratio:
$Br(\kppp) < \newlim\cdot10^{-5}$ at 90\%
confidence level.\\
\vspace*{4mm}
{\it PACS:} 13.25.Es; 11.30.Er; 13.65.+i; 13.75.-n; 14.40.Aq \\
{\it Keywords:} $e^+e^-$ collisions; kaon decays; CP-violation; upper limit
\end{abstract}
\end{singlespace}
\end{frontmatter}

\eject
\section*{\large Introduction}
At present CP-violation is observed only in the
$K_L\to 2\pi$ and $K_L\to\pi l\nu$ decays, and first indication of the effect
in B-decays was recently reported \cite{bdecays}.
Another possible domain for CP-violation studies are still unseeing
$K_S\to\pi^+\pi^-\pi^0$ and $K_S\to 3\pi^0$ decays, of which
the latter must be a pure CP-violating process \cite{ksdecays}, because
for three neutral pions only CP-odd states exist.
CP-violation in the $K_S\to 3\pi^0$ decay can be parameterized
in terms of $\eta_{000}$ parameter, which is defined as:
$$
  \eta_{000} = \frac{A(K_S\to\pi^0\pi^0\pi^0)}{A(K_L\to\pi^0\pi^0\pi^0)}.
$$
One can estimate the decay branching ratio \cite{ksdecays}:
$\mathrm{Br}(\kppp) \simeq |\epsilon_S|^2\cdot
(\tau_S/\tau_L)\cdot\mathrm{Br}(K_L \to 3\pi^0) \sim 10^{-9}$.
The lowest existing experimental upper limit of
$\mathrm{Br}(\kppp) < 1.9\cdot10^{-5}$
was reported by CPLEAR collaboration \cite{CPLEAR}.
In this paper results of the study of the $\kppp$ decay with the SND
detector are presented.

\section*{\large Detector and experiment}
   The experiment was performed in 1996--1998 at VEPP-2M collider
with SND detector \cite{exper}. The SND is a general purpose nonmagnetic
detector. Its main part
is a spherical electromagnetic calorimeter, consisting of 1632
NaI(Tl) crystals.
The calorimeter energy resolution for photons is
$\sigma_E/E=4.2\%/\sqrt[4]{E(\mathrm{GeV})}$,
the angular resolution is
$\sigma_\phi(\mathrm{degrees})=0.82/\sqrt{E(\mathrm{GeV})} \oplus 0.63$
and solid angle coverage is close to $90\%$ of $4\pi$ steradian \cite{detector}.

  Presented analysis is based on experimental data collected in the
center-of-mass energy region
980--1040~$\mathrm{MeV}$, with most of the data,
taken in the close vicinity of the $\phi(1020)$ peak.
The $\pksl$ decays were used as a source of $K_S$ mesons.
Experimental data corresponds to about 
$2\cdot10^7$ produced $\phi$ mesons or $7\cdot10^6$ $K_SK_L$ decays.

\section*{\large Event selection}
  The search for the $\kppp$ decay was performed using the process
\begin{equation}
 e^+e^-\to\phi(1020)\to K_SK_L,\quad \kppp\to6\gamma \label{ks3eq1}.
\end{equation}
In this process the $K_L$ having momentum of about
$110~\mathrm{MeV}$ and decay length of $3.4~\mbox{m}$ may either produce 
signals  in the detector due to nuclear interaction in the calorimeter or
decay in flight, or it can punch through the detector unseen.
The detection efficiency of $K_L$ mesons in the calorimeter material was
studied in the process
\begin{equation}
  e^+e^-\to\phi(1020)\to K_SK_L,\quad \kpp\to4\gamma. \label{ks3eq2} 
\end{equation}
With the probability of $46\%$ the $K_L$ produces a single
cluster of hit crystals
in the calorimeter, and with the $29\%$ probability --- more than one cluster.
The clusters produced by $K_L$ mesons are interpreted as ``photons'' by event
reconstruction program.
The rest $25\%$ of $K_L$ mesons
produce no signal in the calorimeter.

Events with 6 or 7 reconstructed photons were used in the search for
$\kppp$ decay. In order to reject background caused by stray particles from
the accelerator and cosmic events, constraints were imposed on
total energy deposition ($E_\mathrm{tot}>0.35 E_0$) and total momentum of
events ($P_\mathrm{tot}<0.45 E_0$), where $E_0$ is a beam energy.
To suppress cosmic background even further, the events,
where the most of hit crystals could be fitted by a single straight track,
were rejected.
Due to worse energy
resolution near the calorimeter edges the polar angle of all reconstructed
photons
was limited to $30^\circ\le\vartheta\le150^\circ$.

Remaining background comes mainly from two processes:
\begin{equation}
  e^+e^-\to\phi(1020)\to\eta\gamma,\quad \eppp\to 6\gamma \label{ks3eq3}
\end{equation}
and from \rpp\- with $K_L$ 
producing extra 2 or 3 ``photons'' due to nuclear interaction or decay.

Kinematic fitting based on $\chi^2$-method was performed for the events
satisfying selection criteria described above. For each event
two hypotheses were checked:
\begin{itemize}
\item[-] $H_{3\pi}$: an event is due to the process \rppp\-,
 i.e. there are 3 $\pi^0$-s from $K_S$ decay in the event;
\item[-] $H_{2\pi}$: an event is due to the process \rpp\-,
 i.e. it contains 2 $\pi^0$-s from $K_S$ decay.
\end{itemize}
As result of kinematic fitting the following parameters were evaluated:
\begin{itemize}
\item[-] $\chi^2_{3\pi}$ and
$\chi^2_{2\pi}$ --- the chi-square values for the two hypotheses;
\item[-] $P_{3\pi}$ and
$P_{2\pi}$ --- the momentum of reconstructed $K_S$;
\item[-] $\vartheta_{3\pi}$ and
$\vartheta_{2\pi}$ --- the polar angle of reconstructed $K_S$;
\item[-] $m_{3\pi,i}$ and
$m_{2\pi,i}$ --- raw invariant masses of photon pairs, attributed
to pions during kinematic fitting.
\end{itemize}
In order to isolate events of the process \rppp\-,
the following cuts were applied:
\begin{itemize}
\item[ ]$\chi^2_{3\pi}<20, \quad \chi^2_{2\pi}\ge30,$
\item[ ]$80 < P_{3\pi}(\mathrm{MeV}) < 145,$
\item[ ]$30^\circ<\vartheta_{3\pi}<150^\circ.$
\end{itemize}
As a result significant part of background events of the processes \retg\- and
\rpp\- was rejected. The total of
19  6-photon and 15 7-photon events survived the cuts.

\begin{figure}[htbp]
\begin{minipage}[t]{78mm}
\epsfig{figure=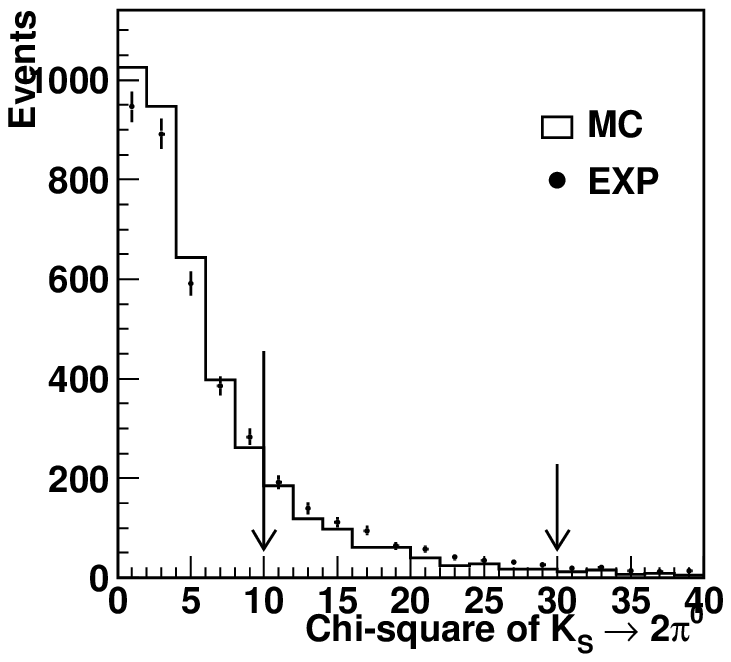}
\caption{\label{ks3fig1}
The distribution of $\chi^2$ of kinematic fit for $\kpp$ events.
The histogram represents the simulation, points --- experimental data.} 
\end{minipage}
\hfill
\begin{minipage}[t]{78mm}
\epsfig{figure=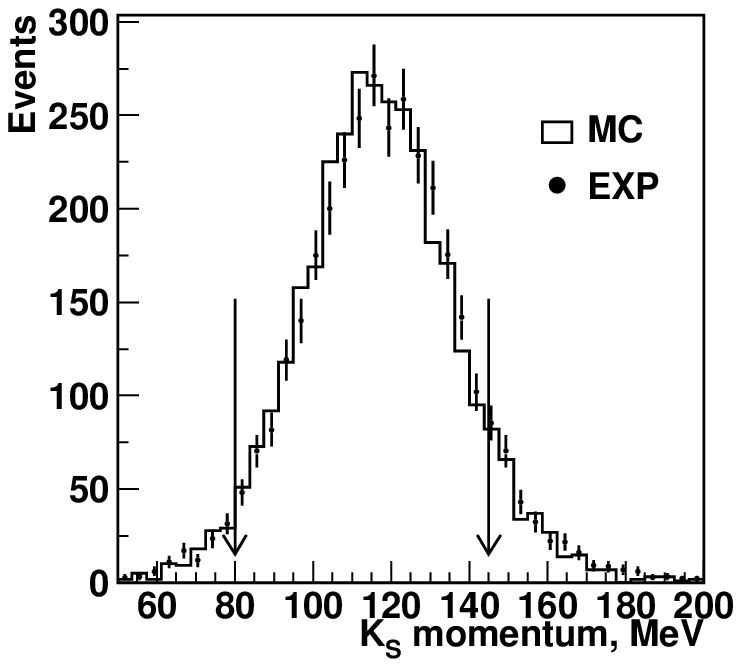}
\caption{\label{ks3fig2}
The distribution of the momentum of reconstructed $K_S$ meson in
$\kpp$ events.
The histogram represents the simulation, points --- data.}
\end{minipage}
\end{figure}

The process \rpp\- was used
as a reference for the detection efficiency monitoring.
The events with 4 and 5 reconstructed photons were selected using the
same primary cuts as for the process \rppp\-.
After kinematic fitting the following additional cuts were applied:
\begin{itemize}
\item[ ]$\chi^2_{2\pi}<10,$ 
\item[ ]$80 < P_{2\pi}(\mathrm{MeV}) < 145,$
\item[ ]$30^\circ<\vartheta_{2\pi}<150^\circ.$
\end{itemize}
 The $\chi^2_{2\pi}$ and $P_{2\pi}$ distributions
for experimental and simulated events of the process \rpp\- are shown in
Fig.~\ref{ks3fig1} and Fig.~\ref{ks3fig2}.

Further analysis of the process \rppp\- candidates was performed
separately for events with 6 and 7 photons.  The reference process \rpp\-
was studied in 4- and 5-photon classes respectively.

\section*{\large Analysis of the events with detected $K_L$ }

\begin{figure}[htbp]
\begin{minipage}[t]{78mm}
\epsfig{figure=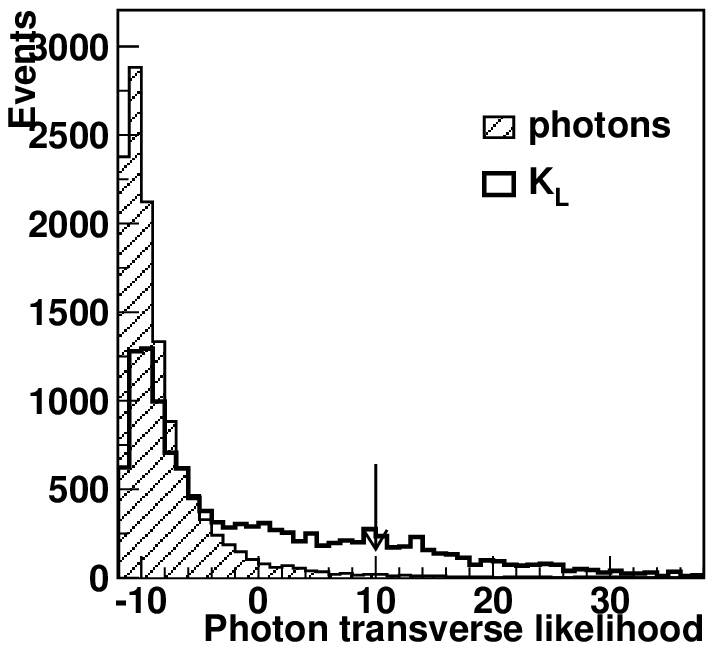}
\caption{\label{ks3fig3}
The $\xi_T$ distributions for photons (clear histogram) and
for $K_L$ mesons (shaded histogram) in the process \rpp\-.}
\end{minipage}
\hfill
\begin{minipage}[t]{78mm}
\epsfig{figure=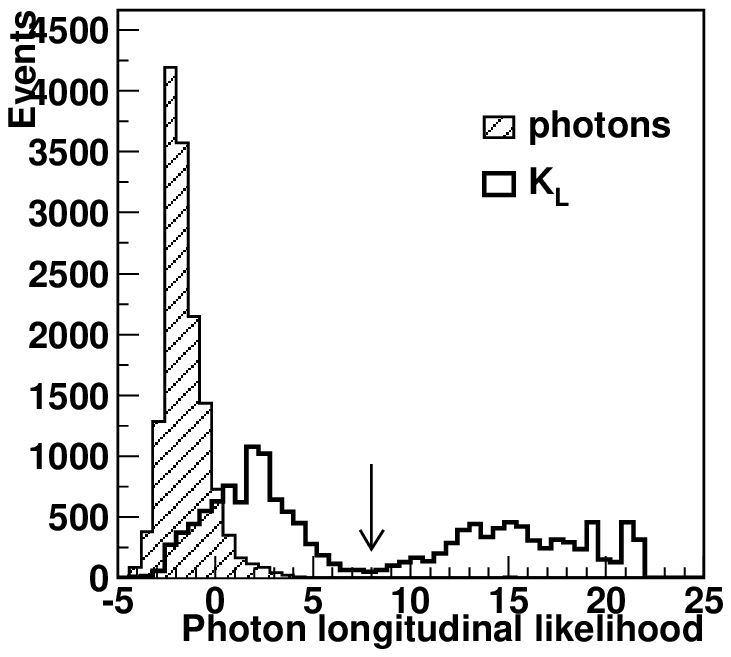}
\caption{\label{ks3fig4}
The $\xi_L$ distributions for photons (clear histogram) and
$K_L$ mesons (shaded histogram) in the process \rpp\-.}
\end{minipage}
\end{figure}

In the events of the process \rppp\- with 7 reconstructed photons, one of the
photons must be in fact a $K_L$ meson, thus additional cuts on its parameters
can be imposed.
In the analysis it was required that energy deposition of
$K_L$ cluster is at least $100~\mathrm{MeV}$, and spatial
angle between the cluster and reconstructed $K_S$ direction
is more than $120$ degrees. The following cuts were based on specific for $K_L$
profiles of energy deposition in the calorimeter.  The parameters $\xi_T$ and
$\xi_L$ were introduced to quantitatively describe the differences between
the energy deposition profiles
for photons and $K_L$ mesons.  The parameter $\xi_T$ \cite{xinm} represents the
likelihood of a hypothesis that the transverse profile of energy deposition
in a cluster was produced by a single photon.  The parameter $\xi_L$ has the
same meaning, but for the longitudinal profile.
Both parameters were studied in
the process \rpp\-, their distributions are shown
in Figs.~\ref{ks3fig3} and \ref{ks3fig4}.
The figures show that the requirement of either
$\xi_T>10$ or $\xi_L>8$ reliably identifies $K_L$ meson.
The same requirement was applied to
$K_L$ meson in 5-photon events of the reference process \rpp\-.

As a result the number of selected events was $N^1_{\kppp} = 0$
for the process \rppp\- and $N^1_{\kpp} = 92676$ for the process \rpp\-.
The detection efficiencies $\varepsilon^1_{\kppp} = 1.7\%$ and
$\varepsilon^1_{\kpp} = 5.3\%$ for the processes \rppp\- and \rpp\- 
were calculated by Monte Carlo simulation using UNIMOD2 package \cite{uni}.
The branching ratio of the $\kppp$ decay was calculated as
follows:
\begin{equation}
 \mathrm{Br}(\kppp) = \mathrm{Br}(\kpp) \cdot
  \frac{ N^1_{\kppp} }{ N^1_{\kpp} } \cdot
  \frac{ \varepsilon^{1}_{\kpp} }{ \varepsilon^{1}_{\kppp}} .
\label{ks3eq5}
\end{equation}
An upper limit was obtained at $90\%$ confidence level:
\begin{equation}
  \mathrm{Br}(\kppp) < 2.4\cdot10^{-5}. \nonumber
\end{equation}

\section*{The analysis of events with undetected $K_L$}

In 6-photon  \rppp\- candidates all detected particle must be photons.
Thus the remaining \rpp\- background can be suppressed by the following
cuts on $\xi_T$ and $\xi_L$:
$\xi_T<0$ and $\xi_L<0$ for all six particles.
It was required also that raw invariant masses of photon pairs, reconstructed
as pions are restricted to the range
$120<m_{3\pi,i}(\mathrm{MeV})<155$.
The same cuts were used in parallel analysis of 4-photon events of the
reference process \rpp\-.
The detection efficiencies of processes \rppp\- and \rpp\-
were obtained by Monte Carlo simulation to be
$\varepsilon^0_{\kppp} = 1.9\%$ and $\varepsilon^0_{\kpp} = 4.3\%$
respectively.  $N^0_{\kppp}=0$ events of the process \rppp\- and
$N^0_{\kpp}=57742$ events of process \rpp\- survived the cuts.
The upper limit at the confidence level of
$90\%$ is:
\begin{equation}
  \mathrm{Br}(\kppp) < 2.8\cdot10^{-5}. \nonumber
\end{equation}

\section*{Combined analysis}
By using the following relation between numbers of found events of the
processes \rppp\- and \rpp\-:
\begin{equation}
 N^{i}_{\kppp}= \mathrm{Br}(\kppp) \cdot
     \frac{ N^{i}_{\kpp}/\varepsilon^i_{\kpp} }{\mathrm{Br}(\kpp)} \cdot
		           \varepsilon^i_{\kppp},
\end{equation}
the results of both analyses can be combined:
\begin{equation}
 \mathrm{Br}(\kppp) = \mathrm{Br}(\kpp) \cdot
    \frac{ (N^{0}_{\kppp}+N^1_{\kppp}) }{
           N^{0}_{\kpp}\frac{\varepsilon^0_{\kppp}}{
	   \varepsilon^0_{\kpp}} +
	   N^{1}_{\kpp}\frac{\varepsilon^1_{\kppp}}{
	   \varepsilon^1_{\kpp}}} \label{ks3eq9} .
\end{equation}
The resulting upper limit according to Eq.(\ref{ks3eq9}) amounts to 
\begin{equation}
  \mathrm{Br}(\kppp) < 1.3\cdot10^{-5} \nonumber
\end{equation}
at the confidence level of $90\%$.

The systematic error of the detection efficiency
is determined mainly by imprecise simulation of $K_L$ nuclear interaction.
Its estimated value is $25\%$.
Since the \rppp\- branching ratio depends on the ratio of the detection
efficiencies for the processes \rppp\- and \rpp\-,
the common systematic error in the $K_L$ simulation cancels.
The remaining systematic error in the efficiency ratio is determined mainly
by the accuracy of simulation of electromagnetic showers.  In order to
estimate this error, the process \retg\- was studied with the cuts
similar to those in the analysis of process \rppp\-.  Resulting branching
ratio of $\phi\to\eta\gamma$ is $8\%$ lower than its world averaged value
\cite{pdg}.
This difference was taken as an estimate of the systematic error
of the ratio of the detection efficiencies.
The final result for the upper limit is then:
\begin{equation}
  \mathrm{Br}(\kppp) < \newlim\cdot10^{-5}. \nonumber
\end{equation}

\section*{Conclusion}

The experiment was performed with SND detector at VEPP-2M $e^+e^-$ collider.
The total statistics of $2\cdot10^7$ $\phi$ mesons was analyzed.
As a result, no candidate events of the $\kppp$  decay were found.
The upper limit of the branching ratio of $\kppp$
$\mathrm{Br}(\kppp) < \newlim\cdot10^{-5}$
at the confidence level of $90\%$ was placed.

\section*{Acknowledgement}
This work is supported in part by The Russian Fund for Basic Researches
(grant 96-15-96327) and STP ``Integration'' (No.274).

\section*{Figure captions}
\begin{enumerate}
\item The distribution of $\chi^2$ of kinematic fit for $\kpp$ events.
The histogram represents the simulation, points --- experimental data.
\item The distribution of the momentum of reconstructed $K_S$ meson in
$\kpp$ events.
The histogram represents the simulation, points --- data.
\item The $\xi_T$ distributions for photons (clear histogram) and
for $K_L$ mesons (shaded histogram) in the process \rpp\-.
\item The $\xi_L$ distributions for photons (clear histogram) and
$K_L$ mesons (shaded histogram) in the process \rpp\-.
\end{enumerate}


\end{document}